\documentclass{article}
\usepackage[utf8]{inputenc}
\usepackage{natbib}
\begin{document}
\title{Cultural Evolution of Categorization}
\author{Pablo Andrés Contreras Kallens$^{1, 2}$, Rick Dale$^{3}$, Paul E. Smaldino$^{1,*}$}
\date{}
\maketitle
\noindent 
$^1$Cognitive and Information Sciences, University of California, Merced, USA\\ 
$^2$Department of Psychology, Cornell University \\
$^3$Department of Communication, University of California, Los Angeles, USA   \\
$^*$ Corresponding author: paul.smaldino@gmail.com \\

\abstract{
Categorization is a fundamental function of minds, with wide ranging implications for the rest of the cognitive system. In humans, categories are shared and communicated {\em between} minds, thus requiring explanations at the population level.
In this paper,  we discuss the current state of research on the cultural evolution of categorization. We begin by delineating key properties of categories in need of evolutionary explanation. We then review computational modeling and laboratory studies of category evolution, including their major insights and limitations. Finally, we discuss remaining challenges for understanding the cultural evolution of categorization.  
}

\section{Introduction}
 
Categorization is a core cognitive skill, with wide-ranging implications for the rest of the cognitive system. Categories allow us to parse our interactions with the world, and divide complex and otherwise chaotic stimuli into discrete kinds. 
Thus, an individual furry moving thing becomes an instance of the category \emph{cat}, which in turn allows us to reason that it would be a bad idea to tug on its tail. In such instances of categorization, perceptual information is compressed and classified on the basis of its relationship to some other previously perceived or conceived category, contextualizing the new stimulus by virtue of similarity or analogy to contexts stored in memory \citep{french1995subtlety,hofstadter2013surfaces}. Novel properties of the present stimuli can be used to adjust or expand existing categories as well as to create new categories.

Although categorization is widely studied in humans, it is a cognitive ability that is necessarily widespread across the animal kingdom, since adaptive decision making is enhanced for those individuals who can best make sense of the world around them. A fascinating example comes from the California ground squirrel, who in the presence of predatory snakes often exhibit “tail flagging,” in which the squirrel rears up and waves its tail vigorously \citep{rundus2007ground}. This prompts the snake to shift its own behavior from predatory to defensive. Ground squirrels not only distinguish snakes from other stimuli, but also notice the difference between rattlesnakes, which are sensitive to infrared signals, and gopher snakes, which are not. In the presence of the former but not the latter predator, ground squirrels will pump blood into their tails while flagging, a costly behavior that is detectable by rattlesnakes but not gopher snakes.

However, a major consideration for the study of categorization in {\em humans} is that humans are profoundly cultural creatures. We show unusually high levels of cooperation and communication, at a degree of complexity that is not seen in other species. Some non-human species are social, and are able to communicate information in ways that reflect the use of elementary ``cultural categories." The alarm call system used by vervet monkeys \citep{seyfarth1980vervet} is a well-known example. Nevertheless, few if any non-human species have {\em cumulative} culture, in which technological and institutional forms can build on innovations from prior or concurrent generations \citep{boyd1996culture,tennie2009ratcheting,legare2017cumulative}. For we humans, culture pervades cognitive and social experience, and thus the cultural nature of our categories is more entrenched. We use categories for talking amongst ourselves about the world, and also for talking about ourselves and others. Communication and coordination in humans require convergence on shared concepts that facilitate common goals, joint attention, and consistent norms and institutions \citep{chwe2001rational,clark1996using,skyrms2004stag,tomasello2009we,smaldino2014cultural}. Our categories must therefore not only be internally consistent; they must be culturally consistent.

Language is often seen as a vehicle for this kind of parcellation of a complex and variegated world. Categorical terms like ``cat'' are a big part of the lexicon of each individual language. Different languages carve up the world in different manners, with some stunning differences in domains such as spatial categories \citep{levinson1994kant,majid2004can}, color \citep{kay2009world}, and kinship \citep{kemp_kinship_2012}. For this reason, language is a critical human skill, subject to functional considerations--such as how a set of semantic categories can help a linguistic community navigate and survive in its environment. Indeed, language is widely discussed in the realms of biological and cultural evolution \citep{christiansen2003language,fitch_evolution_2010,hurford_origins_2012}. Categorization, surprisingly, is less often integrated in evolutionary questions, despite its core cognitive significance. 

Although the cultural features of categories may ultimately manifest in language, there are many important social domains for which the cultural aspects of categorization are paramount. Some examples include:
\begin{itemize}
\item {\em Social identity.} Humans readily classify people into groups and roles that facilitate decision making related to interaction. Such categories are imperative for coordination and assortment in cooperative and competitive tasks \citep{smaldino2018social}, including evaluating potential mates \citep{miller1998mate}.
\item {\em Morality.} Categorizing behaviors as moral, ethical, or legal facilitates a convergence on norms that is necessary for cultural cohesion \citep{curry2016morality}.
\item {\em Emotion.} Interpretations of emotional signals are often bound to context and the culturally appropriate expressions for those contexts \citep{barrett2007language}.  
\item {\em Personality.} The way we describe others relies on finding regularities among behaviors and contexts that are useful for prediction, and these regularities are at least partly culturally determined \citep{gurven2013universal,lukaszewski2017explains}. 
\end{itemize}

Reliance on the categories in the preceding list did not arise spontaneously with the emergence large brains during the course of our evolution. Nor did they emerge through individual learning processes as a result of a shared environment. Many salient categories in human cultures must necessarily arise through cultural processes. And because cultures last far longer than individual lifespans and change through well studied mechanisms \citep{boyd1985culture,turchin2003historical,mesoudi2011cultural}, many properties of human categories and categorization require explanations in terms of cultural evolution.  

In this paper, we discuss the current state of research in understanding the cultural evolution of categorization.  We begin by delineating key properties of categories in need of evolutionary explanation. We then review computational modeling and laboratory studies of category evolution, including major insights and limitations. 
Most of the approaches that we review focus on the cultural evolution of \textit{category terms} and their relationship to the environment. Because of this focus on category terms, prior studies also reflect categorization itself. We will also argue that understanding category systems as the locus of evolutionary dynamics may inform the study of language evolution in interesting ways. More broadly, taking categorization as central may help to integrate various aspects of the debate on cognitive evolution. We will conclude with some remaining challenges for understanding the cultural evolution of categorization.

\section{Features of Categories and Concepts}

For the purpose of evaluating the approaches to the evolution of categories, it is useful to review some of the relevant features of categorization, concepts and their relationship to language. Because of how vast the field of research in concepts and categories is, we will focus on features most relevant to our present goal. Reviews on categorization abound (for authoritative reviews see \cite{murphy2004big,cohen_handbook_2005,pothos_formal_2011}), but some of its core properties can be summarized succinctly. 

Terminologically, we use the term ``concept'' to refer to the \textit{psychological representation} of a category, which allows for the categorization of stimuli. We use the terms ``categories'' or ``categorical structure'' to refer to the partitions of the world that are relatively stable and distributed widely in a population \citep{ross2010concepts}. The latter is encoded in the population's categorical language, shared knowledge structure, and institutions. Categories and concepts are not independent---shared cultural categories may influence individuals' concepts \citep{ross2003cultural}, and of course shared cultural categories depend on the internalized concepts of the constituent individuals . Some aspects of our shared categorical structure are even subject to persistent deliberate change, as with legal or scientific categories. Thus, although our focus will be on culturally shared categories, we cannot avoid some discussion of concepts at the level of individuals. \\

\noindent \emph{\textbf{Categorical boundaries are fuzzy}}. One classical model of categories treats them as a set necessary and sufficient features that uniquely define the set of all their members \citep{murphy2004big}. Empirical psychology has shown this classical thesis to be too simplistic. In their well-known behavioral experiments, Rosch and Mervis (\citeyear{rosch1975family}) showed that categorization is affected by \textit{prototyping effects}. Subjects are able to rank members of a category according to the extent to which they are representative of the category. Most people recognize that apples are better exemplars of fruits than tomatoes. It is hard for the classical view to explain this without further modifications (see Medin \& Smith \citeyear{medin1984concepts}), as apples and tomatoes would be members of the same set by virtue of the very same features. Thus, categorical boundaries are fuzzy in the sense that membership in a category may be relative rather than absolute. 

A related sense in which categories are fuzzy is that they don't appear to have clear-cut boundaries \citep{Rosch1978principles}. Consider the arguments that might ensue when classifying  spiders or fish as pets or non-pets. Ambiguity in category membership can arise even for seemingly clear-cut cases like whether a number is odd or even \citep{armstrong1983some}. Lupyan (\citeyear{lupyan2017paradox}) provides another example using the case of ``triangle.'' Despite its precise, mathematical definition, reasoning about triangles appears to be influenced by secondary descriptors and contexts. This work shows that even highly abstract and formal concepts can have fuzzy boundaries, and points to the role that language plays in making conceptual boundaries more precise.

All categorical terms, at least to some degree, serve to reduce the dimensionality of information. Yet complex stimuli can be decomposed in many different ways. How then should the categories that carve up the world be organized? A version of this question was famously given by Quine (\citeyear{quine2013word}), writing at a time that preceded the tidal wave of experimental psychology on categories. If you heard the word ``gavagai'' in an unknown language, but knew the word was used in the context of a rabbit, you still face considerable challenges of categorization. You would not automatically know if it referred to its color, to the whole rabbit, to a part of the rabbit, or to a broader context like, say, rabbit-on-Tuesday. This does not mean, however, that categories are arbitrary or unconstrained; it just points that there is nothing \textit{in the stimuli} that necessarily suggest a specific way of categorizing, and that our specific set of categories must have developed through some processes in addition to the individual perceiving and interacting with his or her world. Culture, and particularly language, seems to be the main vehicle for the creation of precise categories by a population, and for the acquisition  of less fuzzy concepts by individuals. \\

\noindent \emph{\textbf{Concepts are learned}}. Although it has been argued that some concepts are innate \citep{carey2009origin,wellman_cognitive_1992}, the vast majority of the categories we employ on a day-to-day basis are learned. Laboratory studies have shown that new concepts can be learned by interacting with objects in novel ways \citep{markman2003category}, through generalization of the features of their members \citep{erickson1998rules}, or by association \citep{colunga2005lexicon}. Indeed, decades of experiments have shown that categories can be taught \citep{rosch1975family,medin1978context}. Concepts can be acquired, forgotten, or changed, and can shrink or expand over the course of a lifetime. As a critical example, the vast array of technological artifacts that surround most modern humans have only recently emerged in our history; and some categories that were prominent in the past have fallen out of use. Before the agricultural revolution of 11 thousand years ago, humans lived in predominantly egalitarian societies without extensive hierarchy, division of labor, and wealth of socioeconomic niches that structure lives in contemporary industrialized societies \citep{boehm2009hierarchy,turchin2015ultrasociety}. Thus, not only concepts, but also shared categorical structure can emerge and change over the history of a population. \\

\noindent \emph{\textbf{Categories and concepts are flexible, and goal-oriented}}. Human concepts have several  semantic properties different from those used by non-human animals. Specifically, although most animals are capable of basic categorizing and signaling \citep{fedurek2011primate}, animal calls are typically only \textit{functionally referential} in that they are not flexible and do not seem to respond to generalized discretization of stimuli.\footnote{Although see \cite{fischer1998barbary} and \cite{seyfarth2017origin} for some opposing views.} In contrast, human categorical structure not only draws boundaries between objects, features of objects, processes, etc., but can be used to refer to these categories flexibly and combinatorially. Thus, humans are able to say that although \textit{apes}, \textit{dogs} and \textit{swordfish} are all part of the category \textit{animals}, they have a vast set of features that also make each of them distinct, from their form to their habitat. A dog can be an animal, a pet, a friend, a police officer, or ``something soft.''This property of flexibility makes categorical terms deeply transformative of human intelligence. 

Some have argued that the flexibility and fluidity of categorization is critically important to human creativity and ought to be given more theoretical attention, especially in view of the role of concepts in achieving particular goals. Schemas, for example, are categories for recurrent scenarios that give structure to decision making \citep{smaldino2012origins}. \cite{barsalou1983ad} discussed the importance of goal-oriented ``ad hoc categories,'' which include items such as ``what to sell at a garage sale'' or ``what to take out of the house in case of a fire.'' Any collective action leading to social change requires building new schemas and ad hoc categories, which in turn shape social life. Flexible, goal-oriented categories are therefore often key players in cultural evolution. \\

\noindent \emph{\textbf{Categories and concepts are interwoven with our words for them}}.  
Classically, the relationship between concepts and their linguistic labels has been seen as one in which pre-existing mental representations are associated with an arbitrary word \citep{lupyan2012what}. However, use of concepts is deeply affected by the presence of linguistic labels, over and above the general influence of language on cognition \citep{lupyan2016language}. Categories that are labeled are easier to learn, and there is evidence that concepts activated via linguistic labels are intrinsically different from seemingly identical concepts activated nonverbally \citep{lupyan2012what,lupyan2015meaningless,sufill2016words}.  

Language evolves, in part from pressures to maximize production and understanding, but also in tandem with the categories that shape human life \citep{kirby2017culture,lupyan2016there}. If humans are the talking ape, then we cannot understand ourselves without understanding language. And because language is interwoven with the categories it uses to communicate, we must understand how those categories evolve. \\

\noindent Categorization in humans is complex. 

No doubt there are biological and ecological pressures that favor the development of particular categories that are (i) easy to learn and use and (ii) handy in certain environments. Considering the multitude of objects and ecosystems humans face, though, it is possible to see what can be meant by the cultural evolution of categorization. In evolutionary terms, categorization has a complex fitness landscape. Categorization schemes may emerge and evolve to handle new territories, or to conquer more effectively one that is already at hand. This must happen rapidly, because the social and ecological environments that require categorization can change must more rapidly than genes can adapt. The evolution of categories must therefore be cultural in nature, both for adaptive purposes and for those of social coordination. 

In the remainder of the paper we will review and evaluate a number of modeling and experimental approaches that have been used to study the cultural evolution of categorization. 

\section{Computational models}
Cultural evolutionary processes occur over large spans of space and time, involving complex interactions that are difficult to study directly. Computational models are important tools for understanding such systems \citep{smaldino2017models}. Due both to its centrality in cognitive processing and its utility in commercial software, computational models of categorization are widespread. Indeed, categorization is the primary function of many artificial neural networks since seminal work by \cite{rumelhart1987parallel}, a function which extends to other machine-learning techniques such as K-means clustering \citep{jain2010data}. Our focus, however is on the cultural evolution of categorization. Far less work has been done on this front. This review is intended to be  representative, rather than exhaustive, of the prior modeling work on the social or cultural evolution of categories. We have chosen to emphasize certain works because they focus specifically on the \textit{dynamics} of categorical structure and concepts, be it in the same population or over several generations.

\cite{cangelosi1998emergence} studied a simple model in which agents must forage for mushrooms, and learn to distinguish between two types: edible and poisonous. In some versions of their model, agents can communicate their findings to other agents, who learn to recognize signals representing edible or poisonous and subsequently inform their decisions to advance or withdraw from the location. The categories in this model are provided exogenously, so nothing is really learned about the evolution of categories {\em per se}. However, this study nicely illustrates the idea that the social information can have important consequences for categories. Actively categorizing stimuli is costly and requires time and proximity. Communicating categories can reduce those costs and increase fitness. This model is noteworthy in that categorization serves a clear fitness-enhancing behavior: finding food and avoiding harm. Moreover, in the populations of categorizers that were allowed to communicate without a pre-established vocabulary, a shared structure of categorical terms emerged. Thus, despite its simplicity, this model embraces the goal-oriented and functional nature of categories in a population, and shows how this aspect could be one of the engines driving the creation of categorical structure. As we shall see, most other models ignore such concerns. A tacit assumption of this and other models is that all individuals in the population are cooperative, with aligned interests (e.g., individuals don’t want others to be poisoned). 

Most models of the evolution of categories focus on the evolution and emergence of color concepts \citep{steels2005coordinating,dowman2007explaining,komarova2007evolutionary,puglisi2008cultural,baronchelli2010modeling,gong2010effect,gong2012exploring}. Color categories are a useful case study for at least two reasons. First, there are decades of empirical research on how humans parse colors into categories, including systematic studies of cultural and linguistic differences and universals \citep{kay1999color,kay2009world}. Second, color presents a unidimensional, continuously varying stimuli, so that categories can be straightforwardly modeled as discrete intervals on a finite line segment (although some models have used a circle to model colors; \cite{dowman2007explaining,komarova2007evolutionary}). This is also a limitation of these models in terms of generalizability to other category structures, as we will discuss in a later section. 

Most if not all of these models take as their foundation a paradigm that has come to be known as the {\em Category Game}. The name was first introduced by \cite{puglisi2008cultural}, though it is structurally very similar to paradigms used in earlier papers (e.g. \cite{steels2005coordinating,dowman2007explaining,komarova2007evolutionary}). Before going further, let us contrast this game with the more widely known {\em Signaling Game}, first introduced by \cite{lewis1969convention} and analyzed extensively by \cite{skyrms2010signals} and others. The Signaling Game involves two players: the signaler and the receiver. In its simplest version, the world can be in one of two states. The signaler knows the state of the world, and wants to communicate this to the receiver using one of two signals. The receiver then selects one of two actions, each of which is most appropriate for a particular state of the world. If the receiver selects the appropriate action, both the receiver and the signaler are rewarded. The signals are not inherently meaningful. Rather, it is the task of the individuals to converge on a convention that a particular signals indicates a particular state of the world. It is critical that both the states of the world (the categories) and the signals (the category labels) are fixed, finite sets of equal size. 

In the Category Game model, neither the states of the world (e.g., the color categories) nor the signals (e.g., the color names) are fixed. Instead there is a population of individuals, each of whom keeps a set of labels corresponding to intervals on a color line, such that multiple labels can exist for a given interval. When an interaction occurs, one individual is randomly chosen to be the speaker and the other to be the hearer. The speaker is presented with a set of two or more color swatches, which she must discriminate. She chooses one and names it using either her pre-existing category name or a new name invented to distinguish it from another presented swatch. The hearer then points to the color swatch to which he thinks the color name corresponds. If the hearer fails to correctly identify the correct swatch, he adds the transmitted word to his category discriminating the color interval in question. If there is agreement, the game is a ``success," and both individuals delete from their memories all of the words except the one transmitted from their inventories of words corresponding to the color category in question. The result is generally some sort of population-level consensus on category boundaries and corresponding labels.  

In their extensive study, \cite{steels2005coordinating} modeled each individual with a complex cognitive architecture. Perceptual categorization occurred with an adaptive feedforward neural network, and naming occurred with an associative memory network. Category convergence could occur both through genetic evolution, individual learning, or learning constrained by culture (the need to coordinate on color names). In other words, they were able to test ``nativist," ``empiricist" (or ``evoked culture"), and ``culturist" views on category emergence. They showed that individualist learning (empiricism) resulted in individuals who each had internal categories, but there was not strong coherence among individuals with a population, nor were categories shared across populations. Genetic evolution (nativism) resulted in complete 100\% cohesion of categories within populations, which were not shared across populations. Finally, learning with language led to strong (but not complete) coherence within populations. This is claimed to be somewhat more realistic than the other transmission models both in the extent of the category coherence and in the time frame for that coherence to emerge. In this sense, this model explores the relationship between category learning and communication, and highlights the interaction between individual acquisition of concepts and a culturally evolving shared categorical structure.

\cite{puglisi2008cultural} greatly simplified the model of \cite{steels2005coordinating}. Here, agents perceived color information as a single number and represented category information as explicit look-up tables. They similarly found the emergence of consistent color categories. They also performed computational experiments with different distributions of colors and showed that linguistic categories were more refined in regions where stimuli are more frequent, illustrating how the environment may influence the categorization process. More generally, they found a deviation between many perceptual categories (many color intervals stored in memory) and far fewer linguistic categories. However, this result is likely an artifact of the model assumption that allows for linguistic labels to be eliminated, while there is no mechanisms to eliminate perceptual categories. Considering the evidence for the influence of categorical terms on perceptual categories, it would be reasonable to extend this model to include a process through which perceptual categories can change, for example, when their categorical terms merge or when they fall out of use due to the statistical structure of the environment. Nevertheless, the model is valuable in that it shows how a cultural categorical structure can emerge only constrained by the environment and the need to coordinate with others.

\cite{baronchelli2010modeling} extended this work to demonstrate that the model could generate patterns of color categories consistent with empirical data from the World Color Survey \citep{kay2009world}, a quantitative database of the different ways that over a hundred languages around the world categorize color. In particular, running their model multiple times to represent different cultures, they showed that the resulting dispersion of language terms shared some statistical properties with the empirical data.

Another extension explores the effects of population structure. Gong and colleagues \citep{gong2010effect,gong2012exploring} modeled the Category Game on populations of agents in structured social networks, in contrast with the well-mixed populations used in previous studies. To do this, they introduced two new metrics: {\em overlap}, the average degree of alignment of linguistic categories among individuals (so that a high value indicates individuals tend to develop categories having similar boundaries), and {\em understanding rate}, the percentage of successful category games in all pairs of individuals (so that a high value indicates that individuals tend to use identical labels to describe stimuli with similar boundaries). \cite{gong2012exploring} had agents play the Category Game on several social network structures: well-mixed, ring, 2-D lattice, small-world, scale-free, and star networks. They found that network structure had large effects on population outcomes.  Networks with high average path lengths (such as ring and 2-D lattices) had substantially lower understanding rates and linguistic overlap compared with other network structures. This result was amplified for larger population sizes. Both extensions to the model illustrate a further point about the cultural evolution of categorical structure: it responds to \textit{specific features} of the environment. The structure of a population and the environment with which it interacts affect the particular categories that a population adopts.

\section{Experimental studies}

A related avenue for the exploration of this topic is experimental work. Experiments can be considered as models too, as they are abstract representations of a larger class of social contexts and behavioral opportunities \citep{galantucci2011experimental,schank2014models,smaldino2015theory}. Paradigms used in behavioral experiments are indirectly used to understand different aspects of real evolutionary dynamics. They have several limitations for achieving this goal; most notably, they take place over minutes rather than generations. However, their strength relies on their ability to embed fully realized humans into artificial scenarios. This can be thought of as analogous to the advantages in artificial intelligence research for studying embodied robots instead of simulations: instead of using a simplistic model of the environment, researchers can use the real thing \citep{brooks1991intelligence}. In this case, using real humans instead of computational agents provides data on the kind of cognitive processing humans actually do, even if the environmental setting and evolutionary dynamics are contrived. 

A major experimental paradigm for studying cultural evolution in the laboratory is \textit{iterated learning}, in which participants perform a task and their responses are presented as stimuli for subsequent participants. In the past, these kinds of experiments have been used to study the evolution of language  \citep{kirby2008cumulative,griffiths2007language,galantucci2011experimental,scott2010language}. However, they have also been used to study specific aspects of categorical learning and category term creation. As with our discussion of computational models, our discussion of these studies is meant to be representative rather than exhaustive.

Perfors and Navarro (\citeyear{perfors2014language}) showed that varying the structure of presented stimuli can affect the specific categorical structure of the language that emerges to refer to them. In their study, they presented squares of different sizes and colors to participants paired with randomly generated strings. Each participant was trained in the link between string and square, and then was asked to type the name of unseen squares. The patterns that emerged over time simulated the evolution of a referential linguistic system, and the categories that go along with it. What makes this experiment relevant for this paper is that the authors subjected the participants to the study in three different conditions: an unmodified one; one in which the size of the stimuli varied more discontinuously; and one in which the color of the stimuli varied more discontinuously. This aspect of the experiment simulates differently structured worlds which present different pressures to what kind of categorical system should be created.

As predicted, the languages resulting from the modified conditions had considerable differences. The categories of the emerging languages tended to divide the squares guided by the modified property. Moreover, this effect was gradual over the generations of the chains of participants. Thus, the final categorization patterns were the end result of generations of unknowing cultural evolution interacting with individual learning of concepts.

A recent study by Carr et al. (\citeyear{carr2017cultural}) provides further insight into this process. Their explicit goal was to analyze how chains of participants create categorical terms when not provided with an already categorized world at the beginning of the experiment. As inputs, they created randomly generated triangles with varying edge length. This had the advantage of being a continuous stimulus space, as there is a theoretically infinite number of triangles that can be generated as mid-points between two other triangles. Similarly to the experiment by \cite{perfors2014language} described before, the first generation of participants was trained on a random selection of triangles, accompanied by a randomly generated string. Participants were then asked to generate labels for unseen randomly generated triangles. Starting from the second generation, the training set was the set of triangle-label pairings generated by the person before them in the chain. 

Their first study found that, over successive generations, the number of words present in the languages decreased, as did the transmission errors from one generation to the next. But, more importantly for the goal of this paper, \textit{categorical structure} sharply increased in most of the chains. This structure was measured as the dissimilarity between pairs of strings as they relate to the similarity of the triangles to which they refer \citep{carr2017cultural}. Over time, similar triangles were grouped together by similar, if not identical, words. This result illustrates how the emergence of categorization can partition a continuous space over time through the interaction of individual category learning and cultural transmission. Furthermore, the closeness between the similarity of the stimuli and the terms used sheds light on the link between categorical terms and the use of concepts to categorize objects according to their similarity or dissimilarity.

Though purposefully specific in their scope, both of these experiments reveal the promise of using the iterated learning paradigm to study the evolution of categorization. However, in comparison with the models, these experimental designs tend to omit an essential part of cultural evolution: horizontal transmission. Each generation is simplified as only one person, eliminating the effects that some of the models highlight, such as social coordination. This may facilitate implementation of the design, and subsequent data analysis, but going beyond this simplification is an important next step (e.g., Fay et al., 2010)\nocite{fay2010interactive}. Moreover, the experiments also omit the pragmatic dimension of category use in connection to the emergence of categorical terms. Although more sophisticated than the tasks used in the models, the experimental tasks were still done only for their own sake, with no independent goal upon which to measure how categorization and linguistic labels modified performance over time. Enhancing such deliberate pragmatic dimensions of the categorization tasks would improve the match to the goal-oriented features of categories we described earlier.

\section{Evaluation of Existing Research}

The existing research has provided a crucial foundation for understanding the cultural evolution of categorization and of categories. Nevertheless, we see several promising new paths that may be taken by future research to expand our understanding of this underexplored issue. 

First, models and experiments often lack a fundamental \emph{function} for the categories that emerge. Instead, they are often passive mappings of stimuli. In contrast, the functional quality of natural categories is active, not passive. They reflect things we do to objects, or things that objects might do to us. For this reason, the structure and shape of categories may depend fundamentally on interactive consequences with the world. The lack of function is exemplified by the discrimination task used in the Category Game and related models, which relies solely on coordination of conventions. While coordination is probably a necessary feature for socially shared categories, it is not sufficient in many if not most cases. 

Second, the cultural evolution of categorization must take into account cultural differences. Different cultures exist in diverse ecological settings and have diverse cultural norms that are already known to facilitate crucial differences in cognitive processing \citep{nisbett2001culture,nisbett2005influence,henrich2010weirdest,smaldino2012origins}. The models used to study the cultural evolution and emergence of categories require at their foundation a theory that can take those differences to bear. That said, the study of color terms discussed above it itself promising example of such an analysis. Computational simulation and statistical analysis of cross-cultural tendencies reveals both biological and cultural constraints on these category schemes \citep{regier2007color}. A similar strategy applied more broadly could strengthen the link between cross-cultural category variation and processes of category evolution. This would permit researchers not only to study how categories emerge, but also to explain particular regimes of variation in categories across cultures \citep[e.g.,][]{ross2003cultural}.

Finally, the processes of cultural transmission studied are limited. Cultural evolutionists have long studied the important distinctions between information that is transmitted vertically (from parent to offspring), obliquely (from older non-parent to member of the next generation), and horizontally (between members of the same generation) \citep{cavalli1981cultural,kline2013teaching}. Most models ignore intergenerational evolution, and so consider only horizontal transmission. Meanwhile, most experiments have only considered linear transmission chains that are meant to approximate vertical transmission. Future work might integrate horizontal, vertical, and oblique transmission to more realistically capture potentially critical aspects of how categories emerge. A recent study of language evolution illustrates how this may be possible in experimental studies. \cite{fay2010interactive} developed a laboratory task that could independently manipulate vertical vs. horizontal transmission. They tend to find that horizontal transmission has a surprising impact on the communicability of emergent forms in a language. This kind of design could be adapted for the evolution of category structure. Along these lines, a recent review by \cite{miton2018cumulative} has suggested several additional improvements to cultural transmission experiments, highlighting that horizontal transmission over short time periods may not accurately model vertical transmission over generations. 

Thinking about categories forces us to consider the complexity inherent both out in the world and in our minds. \cite{sperber1996explaining} and others have noted that information transmission is often not well characterized by imitation. Rather, minds reconstruct information based on what they already know, including their other preexisting categories. Thus, the landscape of existing categories is likely to influence future categorization. Evolutionary biologists have studied dynamics like this under the banner of {\em niche construction} \citep{odling2003niche}, and numerous scholars have extended this work to human cultural evolution \citep{ihara2004cultural,kendal2011cultural,laland2000niche,laland2011cultural}. Such work takes note of how individuals actively modify their environments to create new selection pressures, as well as how they inherit the environmental changes made by those who came before them. As such, niche construction may be a useful paradigm for modeling the evolution of categories. Looking outward, categories also emerge through complex social processes involving emergent group structures and iterative interaction over the lifespan \citep{smaldino2014cultural}. Understanding these emergent processes is paramount.

\section{Conclusion}

Previous research has demonstrated that repeated interactions can generate shared categories among members of a population, and that chains of transmission can further shape the landscape of available categories. This research has helped us to understand many of the processes involved in the evolution of categories. An examination of the research as a whole, however, has demonstrated that much work remains for understanding the cultural evolution of categorization. Many of the features of categories described in Section 2 of this paper remain unexplained. Models, be they computational or experimental, are only as good as their assumptions. We have work to do to bring the model assumptions up to speed with what is known about both cognition and cultural transmission. 

Categorization is a fundamental function of minds, and a major topic of research in cognitive science. In humans, however, categories are shared and communicated {\em between} minds, thus requiring an additional level of explanation: the population level. Recently, Cecelia \cite{heyes2018enquire} has made the compelling argument that the fields of cognitive science and cultural evolution have much to offer one another. The problem of understanding the cultural evolution of categorization is perhaps the ideal opportunity for these two fields to make deep contributions and for new synergies to emerge.   

\section*{Acknowledgments}
We thank Karie Moorman and two anonymous reviewers for their helpful comments.

\bibliographystyle{apalike}
\bibliography{references.bib}

\end{document}